# RKDG and Hybrid RKDG+HWENO Schemes for Divergence-Free Magnetohydrodynamics Part I – Formulation and One-Dimensional Tests

By

Dinshaw S. Balsara

(dbalsara@nd.edu)

Department of Physics, University of Notre Dame,

Notre Dame, IN 46556

## Abstract

Balsara [5] showed the importance of divergence-free reconstruction in adaptive mesh refinement problems for magnetohydrodynamics (MHD) and the importance of the same for designing robust second order schemes for MHD was shown in Balsara [7]. In this paper we show that the reconstruction of divergence-free vector fields can be carried out with better than second order accuracy. As a result, we design divergence-free RKDG and hybrid RKDG+HWENO schemes for MHD that have accuracies that are better than second order. Accuracy analysis is carried out and it is shown that the schemes meet their design accuracy for smooth problems. Stringent one-dimensional tests are also presented showing that the schemes perform well on those tests.

# 1 Introduction

The Magnetohydrodynamic (MHD) equations play an important role in many areas of astrophysics, space physics and engineering. As a result, there has been considerable interest in bringing accurate and reliable numerical methods to bear on this problem. The MHD system of equations can be written as a set of hyperbolic conservation laws. As a result, early efforts concentrated on straightforwardly applying second order Godunov techniques to the MHD equations. This was done by Brio and Wu [15], Zachary, Malagoli and Colella [45], Powell [32], Dai and Woodward [22], Ryu and Jones [36], Roe and Balsara [35], Balsara [1], [2] and Falle, Komissarov and Joarder [25]. Recent efforts have focused on understanding the structure of the induction equation:

$$\frac{\partial \mathbf{B}}{\partial t} + \nabla \times (c\,\mathbf{E}) = 0 \qquad (1.1)$$

and the divergence-free evolution that it implies for the magnetic field. In eqn. (1.1), $\mathbf{B}$ is the magnetic field, $\mathbf{E}$ is the electric field and c is the speed of light. The magnetic field starts out divergence-free because of the absence of magnetic monopoles and eqn. (1.1) ensures that it remains divergence-free for all time. The electric field is given by:

$$c\,\mathbf{E} = -\,\mathbf{v} \times \mathbf{B} \qquad (1.2)$$

where $\mathbf{v}$ is the fluid velocity. For the rest of this paper we will simplify the notation by making the transcription $c\,\mathbf{E} \rightarrow \mathbf{E}$. Brackbill and Barnes [12] and Brackbill [13] have shown that violating the $\nabla \cdot \mathbf{B} = 0$ constraint leads to unphysical plasma transport orthogonal to the magnetic field. This comes about because violating the constraint results in the addition of extra source terms in the momentum and energy equations. Yee [44] was the first to formulate divergence-free schemes for electromagnetism. Brecht et al [14] and DeVore [24] did the same for FCT-based MHD. Dai and Woodward [23], Ryu et al [37], Balsara and Spicer [3], Londrillo and DelZanna [30] [31] and Balsara [7] showed that simple extensions of higher order Godunov schemes permit one to formulate divergence-free time-update strategies for the magnetic field. Toth [42] and Balsara and Kim [6] intercompared divergence-cleaning and divergence-free schemes for numerical MHD. The latter authors find that if the test problems are made stringent enough the schemes that are based on divergence-cleaning show significant inadequacies when used for astrophysical applications. Thus it is advantageous to design robust schemes for numerical MHD that are divergence-free, as was done in Balsara [7]. Balsara [7] used the divergence-free reconstruction of vector fields from Balsara [5] to present a formulation that overcame several inconsistencies in previous formulations. It was also shown that the formulation from Balsara [7] was easily extensible to higher order Runge-Kutta Discontinuous Galerkin (RKDG) techniques, and the presentation in that paper is transparently extensible to second order RKDG. Balsara [8] has also formulated hybrid RKDG+Hermite WENO (RKDG+HWENO) schemes for the Euler equations. In such formulations the lower moments of the solution are retained while the higher moments are reconstructed, resulting in low storage schemes that are of great use in large, parallel adaptive mesh refinement (AMR) calculations. The goal of this paper is to present divergence-free RKDG and hybrid RKDG+HWENO formulations for numerical MHD that can be extended to all orders. Because these ideas are very new, we present an exhaustive set of one-dimensional tests, leaving multi-dimensional tests for a subsequent paper.

The need for higher-order schemes for MHD is easy to justify via the following three points. First, Torrilhon [40] has shown that second order schemes for MHD show a pseudo-convergence to unphysical solutions on small and medium-sized meshes and the true convergence to the physical solution only appears on large enough meshes. Torrilhon and Balsara [41] re-examined this problem in light of Weighted Essentially Non-Oscillatory (WENO) schemes and found that such higher order schemes with their large stencils show only a modest improvement. Balsara and Torrilhon [9] showed that RKDG and hybrid RKDG+HWENO formulations for MHD, with their more compact stencils, show a substantial improvement. Second, MHD schemes that are second order accurate show considerable smearing of entropy pulses and Alfven waves. While this can be corrected on a dimension-by-dimension basis using the artificial compression method (ACM) techniques of Yang [43], good treatment in multiple dimensions necessitates higher order schemes. Third, studies of large-scale MHD turbulence produce structure on all scales. Spectral analysis of second order accurate MHD turbulence shows that such structures are strongly damped on a large range of length scales, see Lee et al [29] for a careful detailing of this problem though the basic result had also been known from other studies. This is inevitable because the MHD system admits a larger number of wave families than the Euler system. It is hoped that the more compact stencils of the higher order schemes presented here ameliorate this problem.

Higher order schemes for MHD have been attempted. Jiang and Wu [27] and Balsara and Shu [4] experimented with WENO schemes. Another line of effort stems from the work of Londrillo and DelZanna [31]. As shown by Balsara [7] there is a natural connection between divergence-free reconstruction and RKDG methods. Thus it is useful to make an RKDG formulation for MHD that is explicitly divergence-free in the magnetic field. Discontinuous Galerkin (DG) methods were first introduced by Reed and Hill [34] for solving linear hyperbolic problems associated with neutron transfer. Cockburn and Shu [16], Cockburn, Lin and Shu [17] formulated the method for nonlinear hyperbolic problems and its application to the case of systems of conservation laws in one and multiple dimensions was carried out in Cockburn, Hou and Shu [18] and Cockburn and Shu [19]. Early utilization of RKDG schemes for practical problems involving Euler flow suffered from the fact that scale-free, problem-independent limiters for RKDG had not been formulated. This situation has been rectified in the recent papers by Biswas, Devine and Flaherty [10], Burbeau, Sagaut and Bruneau [11], Krivodonova et al [28], Qiu and Shu [33] and Balsara [8]. Thus RKDG schemes for Euler flows with salient limiting strategies that do not destroy the order of accuracy are well in hand. We build on these advances to formulate divergence-free RKDG and hybrid RKDG+HWENO schemes for MHD in this work using the scale-free problem-independent limiting strategy from Balsara [8].

In Section 2 we catalogue the divergence-free reconstruction of vector fields for higher order schemes. In Section 3 we present the divergence-free formulation of MHD using RKDG methods. In Section 4 we provide an accuracy analysis and in Section 5 we present several test problems.

## 2 Higher Order Divergence-Free Reconstruction of Vector Fields

In this section we study the divergence-free reconstruction of a divergenceless vector field for schemes with better than second order accuracy. In particular, we focus on the third order case. The second order accurate divergence-free reconstruction of vector fields was studied for Cartesian meshes in Balsara [5]. In Balsara [7] we extended this to logically rectangular meshes with diagonal metrics. Balsara [7] also considered the second order accurate divergence-free reconstruction of vector fields on tetrahedral meshes and that too can be extended to higher orders. Since the method was described in detail in Balsara [5] we will focus here on cataloguing results for the higher order case. The reader who wants a pedagogical introduction is referred to Balsara [5] and Balsara [7].

It is important to clarify that several recent works have tried to study numerical electromagnetics using DG schemes, see Hesthaven and Warburton [26] and Cockburn, Li and Shu [21]. Such efforts have focused on using elements with basis functions that are divergence-free within a zone as a way of representing the divergence-free magnetic field. However, such elements produce a discontinuity in the normal component of the magnetic field at the zone boundary. Such elements, nevertheless, seem to be acceptable for certain problems in electromagnetism. As pointed out in Balsara [7], such basis functions would be unacceptable for numerical MHD because the Riemann problem for the physical system becomes ill-defined if there is a jump in the normal component of the magnetic field at a zone face. The fix by Powell [32] can rectify this, but only at the expense of losing momentum and energy conservation. By collocating and limiting the normal components at the zone faces the Balsara [7] and Balsara and Spicer [3] strategies seek to overcome that problem. In the scheme presented in Balsara [7], the normal components of the facially collocated, divergence-free vector field are the fundamental variables. For a reconstruction-based scheme, the moments of each facially-collocated field component are reconstructed within each face. For an RKDG scheme those moments are evolved using a strategy that will be presented in the next section. The divergence-free reconstruction is then a way to use the facial field components and their transverse variation to obtain a representation of the magnetic field everywhere within a zone. Such a reconstruction can be shown to be unique. This is very different from the strategy of Cockburn, Li and Shu [21] where the fundamental variables are coefficients of the basis functions that truly reside within each zone. The reader may feel that this would result in a loss of certain moments, resulting in a decreased order of accuracy. We will show at the end of this section that such is not the case.

For the rest of this work we assume that each zone has local coordinates $(x,y,z) \in [-1/2,1/2] \times [-1/2,1/2] \times [-1/2,1/2]$ . A natural set of modal basis functions within that zone or on its faces would consist of tensor products of the Legendre polynomials $P_0(x)$, $P_1(x)$ and $P_2(x)$ given by:

$$P_0(x) = 1 \; ; \; P_1(x) = x \; ; \; P_2(x) = x^2 - \frac{1}{12} \tag{2.1}$$

The above Legendre polynomials have just been suitably scaled to the local coordinates of the zone being considered. The x-component of the magnetic field in the upper and lower x-faces of this zone can be projected into these bases as:

$$\begin{aligned} B_x(x = \pm 1/2, y, z, t) &= B_0^{x\pm}(t) + B_y^{x\pm}(t) \, y + B_z^{x\pm}(t) \, z \\ &+ B_{yy}^{x\pm}(t) \, (y^2 - 1/12) + B_{yz}^{x\pm}(t) \, y \, z + B_{zz}^{x\pm}(t) \, (z^2 - 1/12) \end{aligned} \tag{2.2}$$

Here $B_0^{x\pm}(t)$ , $B_y^{x\pm}(t)$ , $B_z^{x\pm}(t)$ , $B_{yy}^{x\pm}(t)$ , $B_{yz}^{x\pm}(t)$ and $B_{zz}^{x\pm}(t)$ are the moments of the third order accurate representation in the basis functions that we have chosen. These are the fundamental variables that we will evolve and the evolutionary equations will be developed in the next section. For the rest of this section we drop the time dependence in order to achieve a more compact notation in the reconstruction formulae that we develop. We can write similar expressions for the y and z-components of the field in the appropriate zone faces as:

$$B_y(x, y = \pm 1/2, z) = B_0^{y\pm} + B_x^{y\pm} x + B_z^{y\pm} z \\ + B_{xx}^{y\pm}(x^2 - 1/12) + B_{xz}^{y\pm} x z + B_{zz}^{y\pm}(z^2 - 1/12)$$

(2.3)

$$B_z(x, y, z = \pm 1/2) = B_0^{z\pm} + B_x^{z\pm} x + B_y^{z\pm} y \\ + B_{xx}^{z\pm}(x^2 - 1/12) + B_{xy}^{z\pm} x y + B_{yy}^{z\pm}(y^2 - 1/12)$$

(2.4)

To reconstruct the field in the interior of the zone we pick the following functional forms for the fields:

$$B_x(x, y, z) = a_0 + a_x x + a_y y + a_z z + a_{xx}(x^2 - 1/12) + a_{xy} xy + a_{xz} xz \\ + a_{yy}(y^2 - 1/12) + a_{xyy} x (y^2 - 1/12) + a_{zz}(z^2 - 1/12) + a_{xzz} x (z^2 - 1/12) + a_{yz} yz + a_{xyz} xyz \\ + a_{xxx}(x^3 - 3x/20) + a_{xxy}(x^2 - 1/12) y + a_{xxz}(x^2 - 1/12) z$$

(2.5)

$$B_y(x, y, z) = b_0 + b_x x + b_y y + b_z z + b_{yy}(y^2 - 1/12) + b_{xy} xy + b_{yz} yz \\ + b_{xx}(x^2 - 1/12) + b_{xxy}(x^2 - 1/12) y + b_{zz}(z^2 - 1/12) + b_{yzz} y (z^2 - 1/12) + b_{xz} xz + b_{xyz} xyz \\ + b_{yyy}(y^3 - 3y/20) + b_{xyy} x (y^2 - 1/12) + b_{yyz}(y^2 - 1/12) z$$

(2.6)

$$B_z(x, y, z) = c_0 + c_x x + c_y y + c_z z + c_{zz}(z^2 - 1/12) + c_{xz} xz + c_{yz} yz \\ + c_{xx}(x^2 - 1/12) + c_{xxz}(x^2 - 1/12) z + c_{yy}(y^2 - 1/12) + c_{yyz}(y^2 - 1/12) z + c_{xy} xy + c_{xyz} xyz \\ + c_{zzz}(z^3 - 3z/20) + c_{xzz} x (z^2 - 1/12) + c_{yzz} y (z^2 - 1/12)$$

(2.7)

The rationale for picking this set of moments follows from Balsara [5]. A slight rearrangement of the functional forms has been made in the previous three equations to cast them in terms of the basis functions. In each of eqns. (2.5) to (2.7) the first line represents the terms that are needed for second order reconstruction, the second line gives the terms that are needed for matching the third order terms at the boundaries and the third line gives the terms that are needed for ensuring a divergence-free vector field. The procedure for enforcing the divergence-free constraint is entirely similar to the one in Balsara [5] and will not be repeated here.

We now provide the formulae for obtaining the coefficients in eqn. (2.5). To obtain the coefficients in eqn. (2.6) make the replacements a → b, b → c, c → a, x → y, y → z and z → x in the formulae below. Similarly, to obtain the coefficients in eqn. (2.7) make the replacements a → c, b → a, c → b, x → z, y → x and z → y. Matching the quadratic terms at the $x = \pm 1/2$ boundaries gives:

$$a_{xyy} = B_{yy}^{x+} - B_{yy}^{x-} \;\;;\; a_{xyz} = B_{yz}^{x+} - B_{yz}^{x-} \;\;;\; a_{xzz} = B_{zz}^{x+} - B_{zz}^{x-} \;\;;$$
$$a_{yy} = \frac{1}{2}\left(B_{yy}^{x+} + B_{yy}^{x-}\right) \;\;;\; a_{yz} = \frac{1}{2}\left(B_{yz}^{x+} + B_{yz}^{x-}\right) \;\;;\; a_{zz} = \frac{1}{2}\left(B_{zz}^{x+} + B_{zz}^{x-}\right) \quad (2.8)$$

After making the analogous terms in eqns. (2.6) and (2.7) we are ready to apply the constraints on the cubic terms in eqns. (2.5) to (2.7). After an SVD minimization of $a_{xxy}$ and $a_{xxz}$ we get:

$$a_{xxx} = -\frac{1}{3}\left(b_{xxy} + c_{xxz}\right) \;\;;\; a_{xxy} = -c_{xyz}/4 \;\;;\; a_{xxz} = -b_{xyz}/4 \quad (2.9)$$

Analogous terms in eqns. (2.6) and (2.7) can now be made. Matching the linear terms at the $x = \pm 1/2$ boundaries gives:

$$a_{xy} = B_y^{x+} - B_y^{x-} \quad ; a_{xz} = B_z^{x+} - B_z^{x-} \quad ;$$

$$a_y = -\frac{1}{6} a_{xxy} + \frac{1}{2}\left(B_y^{x+} + B_y^{x-}\right) \quad ; a_z = -\frac{1}{6} a_{xxz} + \frac{1}{2}\left(B_z^{x+} + B_z^{x-}\right)$$

(2.10)

Analogous terms in eqns. (2.6) and (2.7) can now be made. The constraint applied to the quadratic terms in eqns. (2.5) to (2.7) gives:

$$a_{xx} = -\frac{1}{2}\left(b_{xy} + c_{xz}\right) \tag{2.11}$$

Matching the constant terms at the $x = \pm 1/2$ boundaries gives:

$$a_0 = -\frac{1}{6} a_{xx} + \frac{1}{12} a_{yy} + \frac{1}{12} a_{zz} + \frac{1}{2}\left(B_0^{x+} + B_0^{x-}\right) - \frac{1}{24}\left(B_{yy}^{x+} + B_{yy}^{x-}\right) - \frac{1}{24}\left(B_{zz}^{x+} + B_{zz}^{x-}\right)$$

$$a_x = -\frac{1}{10} a_{xxx} + \left(B_0^{x+} - B_0^{x-}\right)$$

(2.12)

The coefficients are so constructed that they ensure the divergence-free constraint:

$$\left(B_0^{x+} - B_0^{x-}\right) + \left(B_0^{y+} - B_0^{y-}\right) + \left(B_0^{z+} - B_0^{z-}\right) = 0 \tag{2.13}$$

This completes our description of the divergence-free reconstruction.

We make a few observations below:
1) We observe that the normal components of the magnetic field in eqns. (2.2) to (2.4) are indeed third order accurate in the faces. Furthermore, specifying all the moments in eqns. (2.2) to (2.4) uniquely specifies all the coefficients in eqns. (2.5) to (2.7). Eqns. (2.5) to (2.7) contain all the third order terms that one would need in a polynomial expansion. Thus all the third order terms in reconstructing a divergence-free vector field in the interior of a zone are already specified by their third order specification at the boundaries.

The few fourth order terms in eqns. (2.5) to (2.7) only help in matching the magnetic fields exactly to the components at the boundaries.

2) For the third order accurate hybrid RKDG+HWENO scheme one would evolve the zeroth order and linear terms in eqns. (2.2) to (2.4) while reconstructing the quadratic terms at each fractional timestep. In three dimensions this can reduce the storage for the magnetic field terms by a factor of two when compared to the corresponding RKDG scheme. Similarly, the storage for the fluid variables is reduced by a factor of 2.5. The savings increase even more as one goes to fourth order schemes. Thus in three dimensions the fourth order accurate hybrid RKDG+HWENO scheme reduces the storage for the magnetic field by a factor of 3.33 while reducing the storage for the fluid variables by a factor of 5 when compared to the corresponding RKDG scheme.

3) The same transformations that were catalogued in Balsara [7] for treating logically rectangular meshes with diagonal metrics go over transparently for the reconstruction given here.

4) The reconstruction given in this section along with the update that is presented in the next section can also be used to obtain globally divergence-free RKDG schemes for electromagnetics. Such a strategy could be useful in problems that have rapid variation in material properties at material interfaces.

## 3 RKDG Schemes for Divergence-free MHD

The equations of ideal MHD can be cast in a conservative form that is suited for the design of higher order Godunov schemes. In that form they become:

$$\frac{\partial \mathbf{U}}{\partial t} + \frac{\partial \mathbf{F}}{\partial x} + \frac{\partial \mathbf{G}}{\partial y} + \frac{\partial \mathbf{H}}{\partial z} = 0 \qquad (3.1)$$

where **F**, **G** and **H** are the ideal fluxes. Written out explicitly, eqn. (3.1) becomes :

$$\frac{\partial}{\partial t}\begin{pmatrix} \rho \\ \rho v_x \\ \rho v_y \\ \rho v_z \\ \varepsilon \\ B_x \\ B_y \\ B_z \end{pmatrix} + \frac{\partial}{\partial x}\begin{pmatrix} \rho v_x \\ \rho v_x^2 + P + \mathbf{B}^2/8\pi - B_x^2/4\pi \\ \rho v_x v_y - B_x B_y/4\pi \\ \rho v_x v_z - B_x B_z/4\pi \\ (\varepsilon + P + \mathbf{B}^2/8\pi) v_x - B_x(\mathbf{v}\cdot\mathbf{B})/4\pi \\ 0 \\ (v_x B_y - v_y B_x) \\ -(v_z B_x - v_x B_z) \end{pmatrix}$$

$$+ \frac{\partial}{\partial y}\begin{pmatrix} \rho v_y \\ \rho v_x v_y - B_x B_y/4\pi \\ \rho v_y^2 + P + \mathbf{B}^2/8\pi - B_y^2/4\pi \\ \rho v_y v_z - B_y B_z/4\pi \\ (\varepsilon + P + \mathbf{B}^2/8\pi) v_y - B_y(\mathbf{v}\cdot\mathbf{B})/4\pi \\ -(v_x B_y - v_y B_x) \\ 0 \\ (v_y B_z - v_z B_y) \end{pmatrix} + \frac{\partial}{\partial z}\begin{pmatrix} \rho v_z \\ \rho v_x v_z - B_x B_z/4\pi \\ \rho v_y v_z - B_y B_z/4\pi \\ \rho v_z^2 + P + \mathbf{B}^2/8\pi - B_z^2/4\pi \\ (\varepsilon + P + \mathbf{B}^2/8\pi) v_z - B_z(\mathbf{v}\cdot\mathbf{B})/4\pi \\ (v_z B_x - v_x B_z) \\ -(v_y B_z - v_z B_y) \\ 0 \end{pmatrix} = 0$$

(3.2)

where $\rho$ is the density, $v_x$, $v_y$ and $v_z$ are the velocity components, $B_x$, $B_y$ and $B_z$ are the magnetic field components, $\gamma$ is the adiabatic index and $\varepsilon = \rho v^2/2 + P/(\gamma - 1) + \mathbf{B}^2/8\pi$ is the total energy. The equations for the density, momentum density and energy density parallel those in the Euler equations and can be discretized using standard RKDG formulations. While the magnetic fields seem to have a conservation law structure, an examination of the flux vectors show that the equations of MHD obey the following symmetries:

$$F_7 = -G_6, \quad F_8 = -H_6, \quad G_8 = -H_7 \qquad (3.3)$$

These symmetries are also obeyed when any manner of non-ideal terms are introduced and are a fundamental consequence of the induction equation, see eqn. (1.1). Balsara and Spicer [3] realized how to use this dualism between the flux components and the electric fields to build electric fields at zone edges using the properly upwinded Godunov fluxes. Balsara [7] introduced a better way of obtaining the electric fields at zone edges that avoids spatial averaging. The Balsara and Spicer [3] scheme is inherently second order accurate because of the spatial averaging. By overcoming this limitation, the Balsara [7] scheme is easily extended to all orders. Once the electric fields are obtained at requisite collocation points on the zone edges a discrete version of eqn. (1.1) can be built, as shown in Balsara [7]. Balsara [7] also showed that Runge-Kutta time-discretizations could be used for MHD.

We begin our formulation of divergence-free RKDG by positing a finite element space of discontinuous functions $\{v_n\}$ within each zone face $A_n$ with normal vector $\mathbf{n}$. The normal component of the magnetic field at that face is then represented in that space of discontinuous functions. Eqn. (2.2) gives an example of the x-component of the magnetic field that is represented in the faces $x = \pm 1/2$ using a basis set that is formed by tensor products of Legendre polynomials. In eqn. (2.2) we use a modal basis set and the coefficients of the Legendre polynomials are the modes of the x-component of the magnetic field. Eqn. (1.1) is then the equation to be discretized using the test functions $\{v_n\}$. Multiplying eqn. (1.1) by each of the basis functions $v_n$ and using the vector identity $\nabla \times (v_n \mathbf{E}) = (\nabla v_n) \times \mathbf{E} + v_n \nabla \times \mathbf{E}$ yields:

$$\frac{d}{dt} \int_{A_n} (\mathbf{n} \cdot \mathbf{B}) \, v_n \, dA_n + \int_{\partial A_n} v_n \, \mathbf{E} \cdot d\mathbf{l} - \int_{A_n} \mathbf{n} \cdot \left[ (\nabla v_n) \times \mathbf{E} \right] dA_n = 0 \qquad (3.4)$$

The basis functions $v_n$ are usually taken to be the basis functions in which the normal component is represented, i.e the very same basis functions used in eqn. (2.2).

In the rest of this section we take eqn. (2.2) as an example and write out the evolutionary equations for the modes of the x-component of the magnetic field using eqn. (3.4). This is done to give the reader an illustrative example of a third order accurate, RKDG scheme for divergence-free MHD. Thus we take $\mathbf{n}$ to be the unit normal in the x-direction and $v_n$ to be the tensor product basis functions formed by using the Legendre polynomials in eqn. (2.2). As in Section 2 we use the local coordinates. Since we restrict our focus to one of the two $x = \pm 1/2$ faces in eqn. (2.2), we leave the "$\pm$" superscript unspecified. As a result, all the terms in the ensuing six equations also pertain to variables within one of those faces. The evolutionary equation for the zeroth moment from eqn. (2.2) takes the form:

$$\frac{d\,B_0^x(t)}{d\,t} + \left[\int E_z(1/2, z)\,dz - \int E_z(-1/2, z)\,dz\right] \\ - \left[\int E_y(y, 1/2)\,dy - \int E_y(y, -1/2)\,dy\right] = 0 \qquad (3.5)$$

The above equation is simply the discrete representation of eqn. (1.1) and, therefore, guarantees divergence-free evolution of the magnetic field. The evolutionary equations for the first moments from eqn. (2.2) take the form:

$$\frac{1}{12}\frac{d\,B_y^x(t)}{d\,t} + \frac{1}{2}\left[\int E_z(1/2, z)\,dz + \int E_z(-1/2, z)\,dz\right] \\ - \left[\int E_y(y, 1/2)\,y\,dy - \int E_y(y, -1/2)\,y\,dy\right] \\ - \left[\int E_z(y, z)\,dy\,dz\right] = 0 \qquad (3.6)$$

and

$$\frac{1}{12}\frac{d\,B_z^x(t)}{d\,t} + \left[\int E_z(1/2, z)\,z\,dz - \int E_z(-1/2, z)\,z\,dz\right] \\ - \frac{1}{2}\left[\int E_y(y, 1/2)\,dy + \int E_y(y, -1/2)\,dy\right] \\ + \left[\int E_y(y, z)\,dy\,dz\right] = 0 \qquad (3.7)$$

The evolutionary equations for the second moments from eqn. (2.2) take the form:

$$\frac{1}{180}\frac{d B^x_{yy}(t)}{d t} + \frac{1}{6}\left[\int E_z(1/2, z)\, dz - \int E_z(-1/2, z)\, dz\right]$$
$$-\left[\int E_y(y, 1/2)\,(y^2 - 1/12)\, dy - \int E_y(y, -1/2)\,(y^2 - 1/12)\, dy\right] \qquad (3.8)$$
$$- 2\left[\int E_z(y, z)\, y\, dy\, dz\right] = 0$$

and

$$\frac{1}{144}\frac{d B^x_{yz}(t)}{d t} + \frac{1}{2}\left[\int E_z(1/2, z)\, z\, dz + \int E_z(-1/2, z)\, z\, dz\right]$$
$$-\frac{1}{2}\left[\int E_y(y, 1/2)\, y\, dy + \int E_y(y, -1/2)\, y\, dy\right] \qquad (3.9)$$
$$-\left[\int E_z(y, z)\, z\, dy\, dz\right] + \left[\int E_y(y, z)\, y\, dy\, dz\right] = 0$$

and

$$\frac{1}{180}\frac{d B^x_{zz}(t)}{d t} + \left[\int E_z(1/2, z)\,(z^2 - 1/12)\, dz - \int E_z(-1/2, z)\,(z^2 - 1/12)\, dz\right]$$
$$-\frac{1}{6}\left[\int E_y(y, 1/2)\, dy - \int E_y(y, -1/2)\, dy\right] \qquad (3.10)$$
$$+ 2\left[\int E_y(y, z)\, z\, dy\, dz\right] = 0$$

Exactly analogous evolutionary equations can be derived for the moments of the y and z components of the magnetic field. The evolutionary equations for the mass, momentum and energy follow the Euler case and explicit expressions for the first few moments have been given in Cockburn and Shu [19] and Balsara [8]. In keeping with the RKDG philosophy, the line integrals in eqns. (3.5) to (3.10) should be based on using the edge-collocated electric fields using the strategy in Balsara [7]. The area integrals in eqns. (3.5) to (3.10) should use electric fields from Riemann problems that are solved within the face being considered. The line integrals in eqns. (3.5) to (3.10) can be done most efficiently by using the Gauss or Gauss-Lobatto quadrature formulae at the edges and the area integration is best done by using tensor products of the Gauss quadrature formulae. Such quadrature formulae are catalogued in Stroud and Secrest [39]. Since efficient implementation of these algorithms in multiple dimensions will be one of the topics to be addressed in the sequel paper, we will not dwell on details associated with quadrature formulae in this paper. A Runge-Kutta timestepping strategy can be used from Shu and Osher [38]. At each fractional time step in the Runge-Kutta temporal update we use the troubled zone indicator from Balsara [8] along with the limiting strategy from Qiu and Shu [33]. This completes our description of the higher order RKDG schemes for divergence-free MHD.

## 4 Accuracy Analysis

We present an accuracy analysis for the RKDG and hybrid RKDG+HWENO schemes described here. The Courant numbers were set to be 0.9 times the maximum permissible values from Cockburn, Karniadakis and Shu [20]. For each test problem the spatial and temporal accuracy were kept the same. We used the sub-cell based monotonicity preserving (MP) algorithm by Balsara [8] for detecting troubled zones for all the tests presented in this and the next section. In Balsara [8] we provided parameter sets for the sub-cell based MP algorithm that were optimized for the Euler equations. The MHD equations are more stringent and extensive testing has yielded the following optimal parameter sets for the sub-cell based MP algorithm:

p=1 RKDG : $\beta = 1.3$; $\alpha = 0.7$; $\kappa = 2.25$; $\tau = 1.1$.
p=2 RKDG : $\beta = 1.2$; $\alpha = 0.55$; $\kappa = 2.25$; $\tau = 1.1$.           (4.1)
p=3 RKDG : $\beta = 1.0$; $\alpha = 0.5$; $\kappa = 2.25$; $\tau = 1.1$.
p=2 Hybrid RKDG+WENO : $\beta = 1.0$; $\alpha = 0.6$; $\kappa = 2.25$; $\tau = 1.1$.
p=3 Hybrid RKDG+WENO : $\beta = 1.0$; $\alpha = 0.5$; $\kappa = 2.25$; $\tau = 1.1$.

The variables in eqn. (4.1) follow the notation that was established in Section 3 of Balsara [8]. The parameters given above were optimized for use with the (local) Lax-Friedrichs flux, which we use all through this work. The above choice of parameters represents conservatively defined sets of choices. The above parameter set was obtained by optimizing the performance of the resultant schemes on all the one-dimensional test problems presented in Ryu and Jones [36] and Dai and Woodward [23]. For many MHD problems the parameters can assume more relaxed values, like the ones given in Balsara [8]. Since the accuracy analysis for entropy waves has already been catalogued in Balsara [8], we focus on Alfven waves here. For all problems in this section and the next we used a computational domain of [-0.5, 0.5].

## 4.a Alfven Wave with Sine Variation

The torsional Alfven wave used in this section consists of :

$$
\begin{aligned}
&\rho = 1 \, ; \, P = 1 \, ; \, B_x = 1 \, ; \, V_A \equiv B_x / \sqrt{4\pi\rho} \, ; \, v_x = V_A \, ; \, V_\perp \equiv 0.2 \, V_A \, ; \\
&v_y = V_\perp \cos\left[2\pi \left(x - (v_x + V_A)\,t\right)\right] ; \\
&v_z = V_\perp \sin\left[2\pi \left(x - (v_x + V_A)\,t\right)\right] ; \\
&B_y = -V_\perp \sqrt{4\pi\rho} \cos\left[2\pi \left(x - (v_x + V_A)\,t\right)\right] ; \\
&B_z = -V_\perp \sqrt{4\pi\rho} \sin\left[2\pi \left(x - (v_x + V_A)\,t\right)\right] ; \\
&\gamma = 1.4
\end{aligned}
\tag{4.2}
$$

The problem was run with periodic boundaries and stopped at a time of 1.7725 , which corresponds to one periodic passage through the domain. Table 1 shows the error in the y-component of the magnetic field in different norms and on meshes with increasing resolution. The sinusoidal function is very smooth. As a result, we see that the solution with the limiter has the same value as the solution without the limiter for all the second and third order accurate schemes, showing us that the zones were never flagged as troubled. We also see that the solution with the limiter differs from the solution without the limiter for the fourth order accurate scheme. This is because the parameters in the detector for troubled zones have to be set very stringently for the fourth order schemes. If the parameters are not set stringently for the fourth order schemes then they will show some deficiencies on a large enough suite of MHD test problems, like the one mentioned in the previous paragraph. Table 1 shows that all schemes achieve their design accuracies.

## 4.b Alfven Wave with Sine³ Variation

The torsional Alfven wave used in this section consists of :

$$\rho = 1 \;;\; P = 1 \;;\; B_x = 1 \;;\; V_A \equiv B_x / \sqrt{4\pi\rho} \;;\; v_x = V_A \;;\; V_\perp \equiv 0.2\, V_A \;;$$
$$\psi = \sin\left[2\pi\left(x - (v_x + V_A)t\right)\right] \;;\; \chi = \cos\left[2\pi\left(x - (v_x + V_A)t\right)\right] \;;$$
$$v_y = V_\perp \sqrt{1 - \psi^{2m}}\; \text{sgn}(\chi) \;;$$
$$v_z = V_\perp \, |\psi|^m \, \text{sgn}(\psi) \;; \qquad (4.3)$$
$$B_y = -V_\perp \sqrt{4\pi\rho}\sqrt{1-\psi^{2m}}\; \text{sgn}(\chi) \;;$$
$$B_z = -V_\perp \sqrt{4\pi\rho}\, |\psi|^m \, \text{sgn}(\psi) \;;$$
$$\gamma = 1.4 \;;\; m = 3$$

The problem was run with periodic boundaries and stopped at a time of 1.7725, which corresponds to one periodic passage through the domain. The variable "m" can take on increasing values to produce increasingly stringent test problems and we use m=3 here. The higher power in the sine function ensures that higher order derivatives of $v_z$ and $B_z$ are also zero at the zeros of the sine function, yielding a significantly more stringent test. The higher power in the sine function also produces a rapidly varying curvature in the solution, adding to the stringency of the test. Table 2 shows the error in the y-component of the magnetic field in different norms and on meshes with increasing resolution. All schemes can be seen to achieve their design accuracies. For this problem we see that there are small differences between the solution with the limiter and the one without the limiter even in the second order case. For the third and fourth order schemes we see that the difference between the solution with the limiter and the one without is quite significant on coarse meshes. However, as the mesh is refined the solutions assume comparable accuracy. This shows us that on fine enough meshes the third and fourth order schemes almost recover the accuracy of the solution without the limiter. We also see that the p=2 RKDG and the hybrid p=2 RKDG+HWENO schemes have accuracies that do not differ by a lot. This shows us the merit of using p=2 RKDG+HWENO schemes, with their substantially lower storage, in large AMR-MHD applications. The analogous result for the Euler equations was first pointed out in Balsara [8]. The p=3 RKDG and the hybrid p=3 RKDG+HWENO schemes do show a greater difference, an observation that was also made in Balsara [8].

## 5 Test Problems

In this section we present several Riemann problem tests for the various RKDG and hybrid RKDG+HWENO schemes that have been designed here. The ratio of specific heats was set to 5/3 for all the test problems. The left and right states for each of the Riemann problems are given below. 400 zones were used in all tests.

### 5.1 Compound shocks

This co-planar problem, patterned after Brio and Wu [15], consists of the following initial conditions:

$$(\rho_L, P_L, v_{x,L}, v_{y,L}, v_{z,L}, B_{x,L}, B_{y,L}, B_{z,L}) = (1, 1, 0, 0, 0, 0.75\sqrt{4\pi}, 1\sqrt{4\pi}, 0)$$
$$(\rho_R, P_R, v_{x,R}, v_{y,R}, v_{z,R}, B_{x,R}, B_{y,R}, B_{z,R}) = (0.125, 0.1, 0, 0, 0, 0.75\sqrt{4\pi}, -1\sqrt{4\pi}, 0)$$

(5.1)

The problem was run with the second order accurate p=1 RKDG scheme and the results at a time of 0.1 are shown in Fig. 1. The problem has a left-going fast magnetosonic rarefaction, a left-going compound shock, a contact discontinuity, a right-going slow magnetosonic shock and a right-going fast magnetosonic rarefaction. We see that all the shocks are crisp and the contact discontinuity has been captured with a small number of zones. We did not use any special compression algorithms like ACM to restore crispness to the contact discontinuity. Higher order schemes in this family of schemes cause the contact discontinuity to become even better resolved, as we will see in subsequent tests.

## 5.2 Strong Shocks

This planar problem was drawn from Ryu and Jones [36] and consists of the following initial conditions:

$$(\rho_L, P_L, v_{x,L}, v_{y,L}, v_{z,L}, B_{x,L}, B_{y,L}, B_{z,L}) = (1, 20, 10, 0, 0, 5, 5, 0)$$
$$(\rho_R, P_R, v_{x,R}, v_{y,R}, v_{z,R}, B_{x,R}, B_{y,R}, B_{z,R}) = (1, 1, -10, 0, 0, 5, 5, 0)$$

(5.2)

The problem was run with the third order accurate p=2 RKDG scheme and the results at a time of 0.08 are shown in Fig. 2. The problem has a left-going fast magnetosonic shock, a left-going slow magnetosonic rarefaction, a contact discontinuity, a right-going slow magnetosonic shock and a right-going fast magnetosonic shock. We see that all the strong shocks are crisp and the contact discontinuity has been captured with a very small number of zones. The higher order scheme shows its effectiveness by capturing the contact discontinuity without much smear despite the fact that the contact discontinuity is very slow-moving. Second order schemes can be made to capture contact discontinuities crisply, but only with the help of compressive steepeners like the ACM steepener of Yang [43]. The problem with using such steepeners is that they also introduce jitters into the solution in the vicinity of strong shocks. It is difficult to find effective, problem-independent strategies that turn the steepeners on away from shocks and turn them off in the vicinity of strong shocks, but see Balsara [2]. The higher order schemes show their strength by capturing the strong shocks well while also retaining good resolution at linearly degenerate discontinuities. The other third and fourth order accurate schemes formulated here produce results on this test problem that are of comparable quality.

## 5.3 All Seven waves

This non-planar problem was drawn from Ryu and Jones [36] and consists of the following initial conditions:

$$(\rho_L, P_L, v_{x,L}, v_{y,L}, v_{z,L}, B_{x,L}, B_{y,L}, B_{z,L}) =$$
$$(1.08, 0.95, 1.2, 0.01, 0.5, 2, 3.6, 2.0)$$
$$(\rho_R, P_R, v_{x,R}, v_{y,R}, v_{z,R}, B_{x,R}, B_{y,R}, B_{z,R}) =$$
$$(1, 1, 0, 0, 0, 2, 4, 2)$$
(5.3)

The problem was run with the fourth order accurate p=3 RKDG scheme and the results at a time of 0.2 are shown in Fig. 3. The problem has left and right-going magnetosonic shocks of either family, a contact discontinuity and left and right-going torsional Alfven waves. We see that all the shocks are crisp and the contact discontinuity and torsional Alfven waves have been captured with a very small number of zones. The other third and fourth order accurate schemes formulated here produce results on this test problem that are of comparable quality.

## 5.4 Colliding MHD Streams

This co-planar problem was drawn from Dai and Woodward [22] and consists of the following initial conditions:

$$(\rho_L, P_L, v_{x,L}, v_{y,L}, v_{z,L}, B_{x,L}, B_{y,L}, B_{z,L}) =$$
$$(0.15, 0.28, 21.55, 1, 1, 0, -2, -1)$$
$$(\rho_R, P_R, v_{x,R}, v_{y,R}, v_{z,R}, B_{x,R}, B_{y,R}, B_{z,R}) =$$
$$(0.1, 0.1, -26.45, 0, 0, 0, 2, 1)$$
(5.4)

The problem was run with the third order accurate p=2 RKDG+HWENO scheme and the results at a time of 0.04 are shown in Fig. 4. It consists of two high Mach number flows that collide with each other. The problem has left and right-going fast magnetosonic shocks and a contact discontinuity. Despite the shocks being very strong, we see that they are captured without any spurious oscillations. The contact discontinuity is also captured with very few zones, showing the usefulness of a higher order scheme. The other third and fourth order accurate schemes formulated here produce results on this test problem that are of comparable quality.

## 5.5 MHD Analogue of Noh Problem

This planar problem was drawn from Dai and Woodward [22] and consists of the following initial conditions:

$$(\rho_L, P_L, v_{x,L}, v_{y,L}, v_{z,L}, B_{x,L}, B_{y,L}, B_{z,L}) =$$
$$(1, 1, 36.87, -0.155, -0.0386, 4, 4, 1)$$
$$(\rho_R, P_R, v_{x,R}, v_{y,R}, v_{z,R}, B_{x,R}, B_{y,R}, B_{z,R}) =$$
$$(1, 1, -36.87, 0, 0, 4, 4, 1)$$
(5.5)

The problem was run with the fourth order accurate p=3 RKDG+HWENO scheme and the results at a time of 0.03 are shown in Fig. 5. It consists of two extremely high Mach number flows that collide with each other. The problem has left and right-going fast magnetosonic shocks of nearly infinite strength. There is evidence for some wall-heating in the center but it is not too damaging. The spurious central spikes in the transverse velocities were also found in Dai and Woodward [22] and higher order schemes do not seem to cure them. Despite the near-infinite strength of the shocks the fourth order scheme handles the problem well, showing that the methods presented here do effectively combine the dual, and often-conflicting, demands of capturing very strong shocks and retaining low dissipation at linearly degenerate discontinuities. The other third and fourth order accurate schemes formulated here produce results on this test problem that are of comparable quality.

## 6 Conclusions

The work presented here enables us to come to the following conclusions:
1) Following a line of development begun in Balsara [5], we show that the problem of reconstructing divergence-free vector fields can be carried out to higher orders.
2) Following a line of development begun in Balsara [7], we show that the above development yields RKDG and hybrid RKDG+HWENO schemes with order of accuracy that is better than second. In particular, we explore the third and fourth order accurate schemes here.

3) When applied to smooth test problems, the schemes have been shown to meet their design accuracies both with and without limiters.

4) Using a stringent set of test problems we show that the schemes presented here effectively combine the dual, and often-conflicting, demands of capturing very strong shocks and retaining low dissipation in contact discontinuities and Alfven waves. This shows the effectiveness of the schemes for numerical MHD.

## Acknowledgements

DSB acknowledges support via NSF grants R36643-7390002, AST-005569-001 and DMS-0204640.

# References

[1] D.S. Balsara, *Linearized formulation of the Riemann problem for adiabatic and isothermal magnetohydrodynamics*, Astrophysical Journal Supplement, 116 (1998), 119

[2] D.S. Balsara, *Total variation diminishing algorithim for adiabatic and isothermal magnetohydrodynamics,* Astrophysical Journal Supplement, 116 (1998), 133

[3] D.S. Balsara and D.S. Spicer, *A staggered mesh algorithm using higher order Godunov fluxes to ensure solenoidal magnetic fields in MHD simulations*, Journal of Computational Physics, 149 (1999), 270.

[4] D.S. Balsara and C.-W. Shu, *Monotonicity Preserving Weighted Essentially Non-Oscillatory Schemes with Increasingly High Order of Accuracy*, Journal of Computational Physics, 160 (2000), 405.

[5] D.S. Balsara, *Divergence-free adaptive mesh refinement for magnetohydrodynamics*, Journal of Computational Physics, 174(2) (2001), 614-648.

[6] D.S. Balsara and J.S. Kim, *An intercomparison between divergence-cleaning and staggered mesh formulations for numerical MHD*, Astrophysical Journal, 602 (2004) 1079-1090

[7] D.S. Balsara, *Second order accurate schemes for MHD with divergence-free reconstruction*, Astrophysical Journal Supplements, 151(1), (2004) 149-184

[8] D.S. Balsara, *A sub-cell based indicator for troubled zones in RKDG schemes and a novel class of hybrid RKDG+HWENO schemes*, submitted, Journal of Computational Physics, (2004).

[9] D.S. Balsara and M. Torrilhon, *High Order RKDG Schemes: Investigations on Non-Uniform Convergence for MHD Riemann Problems*, in preparation, Journal of Computational Physics, (2004).

[10] R. Biswas, K.D. Devine and J. Flaherty, *Parallel adaptive finite element methods for conservation laws*, Applied Numerical Mathematics, 14 (1994), 255-283.

[11] A. Burbeau, P. Sagaut and C.H. Bruneau, *A problem-independent limiter for high-order Runge-Kutta discontinuous Galerkin methods*, Journal of Computational Physics, 169 (2001), 111-150.

[12] J.U. Brackbill and D.C. Barnes, *The effect of nonzero $\nabla \cdot \mathbf{B}$ on the numerical solution of the MHD equations*, J. Comput. Phys., 35 (1980), 462

[13] J.U. Brackbill, *Fluid modeling of magnetized plasmas*, Space Sci. Rev., 42 (1985) 153

[14] S.H. Brecht, J.G. Lyon, J.A. Fedder, and K. Hain, Geophysical Research Letters, 8 (1981), 397

[15] M. Brio and C.C. Wu, *An upwind differencing scheme for the equations of ideal MHD*, Journal of Computational Physics, 75 (1988), 400

[16] B. Cockburn and C.-W. Shu, *TVB Runge-Kutta local projection discontinuous Galerkin finite element method for conservation laws II: general framework*, Mathematics of Computation, 52 (1989) 411-435.

[17] B. Cockburn, S.-Y. Lin and C.-W. Shu, *TVB Runge-Kutta local projection discontinuous Galerkin finite element method for conservation laws III: one dimensional systems*, Journal of Computational Physics, 84 (1989), 90-113.

[18] B. Cockburn, S. Hou and C.-W. Shu, *TVB Runge-Kutta local projection discontinuous Galerkin finite element method for conservation laws IV: the multidimensional case*, Mathematics of Computation, 54 (1990), 545-581.

[19] B. Cockburn and C.-W. Shu, *TVB Runge-Kutta local projection discontinuous Galerkin finite element method for conservation laws V: multidimensional systems*, Journal of Computational Physics, 141, (1998), 199-224.

[20] B. Cockburn, G. Karniadakis and C.-W. Shu, *The development of discontinuous Galerkin Methods, in Discontinuous Galerkin Methods: Theory, Computation and Applications*, B. Cockburn, G. Karniadakis and C.-W. Shu, editors, Lecture Notes in Computational Science and Engineering, volume 11, Springer, 2000, Part I: Overview, 3-50.

[21] B. Cockburn, F. Li and C.-W. Shu, *Locally divergence-free discontinuous Galerkin methods for the Maxwell equations* , Journal of Computational Physics, 194 (2004) 588-610.

[22] W. Dai and P.R. Woodward, *Extension of the piecewise parabolic method (PPM) to multidimensional MHD*, Journal of Computational Physics, 111 (1994), 354

[23] W. Dai and P.R. Woodward, *On the divergence-free condition and conservation laws in numerical simulations for supersonic MHD flows*, Astrophysical Journal, 494 (1998), 317

[24] C.R. DeVore, *Flux corrected transport techniques for multidimensional compressible MHD*, Journal of Computational Physics, 92 (1991), 142

[25] S.A.E.G. Falle, S.S. Komissarov and P. Joarder, *A multidimensional upwind scheme for MHD*, Monthly Notices of the Royal Astronomical Society, 297 (1998), 265

[26] J.S. Hesthaven and T. Warburton, *Nodal high-order methods on unstructured grids I. Time-domain solution of Maxwell's equations*, Journal of Computational Physics, 181 (2002), 186-221.

[27] G.-S. Jiang and C.C. Wu, *A high-order WENO finite difference scheme for the equations of ideal MHD* , Journal of Computational Physics, 150(2) (1999), 561-594

[28]] L. Krivodonova, J. Xin, J.-F. Remacle and J.E. Flaherty, *Shock detection and limiting with discontinuous Galerkin methods for hyperbolic conservation laws* , Applied Numerical Mathematics, 48 (2004), 323-338.

[29] H. Lee, D. Ryu, J.S. Kim, T.W. Jones, and D.S. Balsara, *Effects of magnetic fields on two-dimensional compressible turbulence*, Astrophysical Journal, 594 (2003), 627

[30] P. Londrillo and L. DelZanna, *High-Order Upwind Schemes for Multidimensional Magnetohydrodynamics*, Astrophysical Journal, 530 (200), 508

[31] P. Londrillo and L. Del Zanna, *On the divergence-free condition in Godunov-type schemes for ideal MHD: the upwind constrained transport method*, Journal of Computational Physics 195 (2004) 17-48.

[32] K.G. Powell, *An Approximate Riemann Solver for MHD ( that actually works in more than one dimension)*, ICASE Report No. 94-24, Langley VA, (1994)

[33] J. Qiu and C.-W. Shu, *Runge-Kutta discontinuous Galerkin method using WENO limiters*, submitted, SIAM Journal on Scientific Computing, (2004a)

[34] W.H. Reed and T.R. Hill, *Triangular mesh methods for neutron transport equation*, Tech. Report LA-UR-73-479, Los Alamos Scientific Laboratory, 1973.

[35] P.L. Roe and D.S. Balsara, *Notes on the Eigensystem for MHD*, SIAM Journal of Applied Mathematics, 56 (1996), 57

[36] D. Ryu and T.W. Jones, *Numerical MHD in Astrophysics: Algorithm and Tests for One-Dimensional Flow*, Astrophysical Journal, 442 (1995), 228

[37] D. Ryu, F. Miniati, T.W. Jones and A. Frank, *A Divergence-free Upwind Code for Multi-dimensional MHD Flows*, Astrophysical Journal, 509 (1998), 244

[38] C.-W. Shu and S. Osher, *Efficient implementation of essentially non-oscillatory shock-capturing schemes*, Journal of Computational Physics, 77, (1988), 439-471.

[39] A.H. Stroud and D. Secrest, *Gaussian Quadrature Formulas*, Prentice-Hall Inc. (1966).

[40] M. Torrilhon, *Non-uniform convergence of finite-volume schemes for Riemann problems of ideal MHD*, Journal of Computational Physics, 69(3) (2003), 253

[41] M. Torrilhon and D.S. Balsara, *High Order WENO Schemes: Investigations on Non-Uniform Convergence for MHD Riemann Problems*, to appear Journal of Computational Physics, (2004).

[42] G. Toth, *The $\nabla \cdot \mathbf{B} = 0$ Constraint in Shock-Capturing MHD Codes*, Journal of Computational Physics, 161 (2000), 605

[43] H. Yang, *An Artificial Compression Method for ENO Schemes, the Slope Modification Method*, Journal of Computational Physics, 89 (1990), 125

[44] K.S. Yee, *Numerical Solution of Initial Boundary Value Problems Involving Maxwell Equation in an Isotropic Media*, IEEE Trans. Antenna Propagation, 14 (1966), 302

[45] A.L. Zachary, A. Malagoli and P. Colella, *A Higher Order Godunov Method for Multidimensional Ideal MHD*, SIAM J. Sci. Comput., 15 (1994), 263

# Tables

Table 1: Alfven wave with sine profile as given in eqn. 4.2. The problem was run on the domain [-0.5, 0.5] with periodic boundaries and stopped at a time of 1.7725. Comparing DG with and without limiter. The MP detection algorithm was used with MHD settings. The error in the y-component of the magnetic field is shown in the $L_1$ and $L_\infty$ norms.

| | N | \multicolumn{4}{c}{DG or DG+HWENO With WENO Limiter} | \multicolumn{4}{c}{DG or DG+HWENO Without Limiter} |
| | | $L_1$ err | order | $L_\infty$ err | order | $L_1$ err | order | $L_\infty$ err | order |
|---|---|---|---|---|---|---|---|---|---|
| $P^1$-RKDG | 20 | 2.06e-4 | | 3.22e-4 | | 2.06e-4 | | 3.22e-4 | |
| | 40 | 4.54e-5 | 2.18 | 7.14e-5 | 2.18 | 4.54e-5 | 2.18 | 7.14e-5 | 2.18 |
| | 80 | 1.09e-5 | 2.07 | 1.71e-5 | 2.07 | 1.09e-5 | 2.07 | 1.71e-5 | 2.07 |
| | 160 | 2.66e-6 | 2.03 | 4.18e-6 | 2.03 | 2.66e-6 | 2.03 | 4.18e-6 | 2.03 |
| $P^2$-RKDG | 20 | 1.12e-6 | | 1.77e-6 | | 1.12e-6 | | 1.77e-6 | |
| | 40 | 7.14e-8 | 3.97 | 1.12e-7 | 3.98 | 7.14e-8 | 3.97 | 1.12e-7 | 3.98 |
| | 80 | 6.55e-9 | 3.45 | 1.03e-8 | 3.45 | 6.55e-9 | 3.45 | 1.03e-8 | 3.45 |
| | 160 | 7.41e-10 | 3.14 | 1.16e-9 | 3.14 | 7.41e-10 | 3.14 | 1.16e-9 | 3.14 |
| $P^3$-RKDG | 20 | 1.71e-6 | | 2.67e-6 | | 1.81e-10 | | 2.84e-10 | |
| | 40 | 1.92e-8 | 6.49 | 2.99e-8 | 6.48 | 2.34e-11 | 2.95 | 3.67e-11 | 2.95 |
| | 80 | 1.07e-9 | 4.17 | 1.67e-9 | 4.17 | 1.49e-12 | 3.97 | 2.33e-12 | 3.97 |
| | 160 | 3.66e-11 | 4.86 | 5.75e-11 | 4.86 | 5.95e-14 | 4.64 | 9.32e-14 | 4.64 |
| $P^2$-RKDG HWENO | 20 | 1.06e-5 | | 1.63e-5 | | 1.06e-5 | | 1.63e-5 | |
| | 40 | 6.75e-7 | 3.96 | 1.06e-6 | 3.94 | 6.75e-7 | 3.96 | 1.06e-6 | 3.94 |
| | 80 | 4.23e-8 | 3.99 | 6.64e-8 | 3.99 | 4.23e-8 | 3.99 | 6.64e-8 | 3.99 |
| | 160 | 2.67e-9 | 3.99 | 4.19e-9 | 3.99 | 2.67e-9 | 3.99 | 4.19e-9 | 3.99 |
| $P^3$-RKDG HWENO | 20 | 3.74e-6 | | 5.80e-6 | | 3.56e-8 | | 5.54e-8 | |
| | 40 | 3.52e-9 | 10.05 | 1.56e-9 | 10.03 | 2.34e-9 | 3.92 | 3.69e-9 | 3.91 |
| | 80 | 9.44e-10 | 1.89 | 1.48e-9 | 1.89 | 9.78e-11 | 4.58 | 1.53e-10 | 4.58 |
| | 160 | 3.57e-11 | 4.72 | 5.60e-11 | 4.72 | 3.23e-12 | 4.92 | 5.07e-12 | 4.92 |

Table 2: Alfven wave with sine³ profile as given in eqn. 4.3. The problem was run on the domain [-0.5, 0.5] with periodic boundaries and stopped at a time of 1.7725. Comparing DG with and without limiter. The MP detection algorithm was used with MHD settings. The error in the y-component of the magnetic field is shown in the $L_1$ and $L_\infty$ norms.

|  |  | DG or DG+HWENO With WENO Limiter |  |  |  | DG or DG+HWENO Without Limiter |  |  |  |
| --- | --- | --- | --- | --- | --- | --- | --- | --- | --- |
|  | N | $L_1$ err | order | $L_\infty$ err | order | $L_1$ err | order | $L_\infty$ err | order |
| $P^1$-RKDG | 40 | 4.72e-4 |  | 1.44e-3 |  | 2.90e-4 |  | 6.31e-4 |  |
|  | 80 | 6.14e-5 | 2.94 | 1.39e-4 | 3.37 | 5.34e-5 | 2.44 | 1.24e-4 | 2.35 |
|  | 160 | 1.29e-5 | 2.25 | 2.93e-5 | 2.25 | 1.25e-5 | 2.09 | 2.92e-5 | 2.08 |
|  | 320 | 3.07e-6 | 2.07 | 7.19e-6 | 2.03 | 3.07e-6 | 2.03 | 7.19e-6 | 2.03 |
| $P^2$-RKDG | 40 | 1.80e-3 |  | 4.52e-3 |  | 5.40e-6 |  | 9.30e-6 |  |
|  | 80 | 8.02e-5 | 4.49 | 4.27e-4 | 3.40 | 2.97e-7 | 4.19 | 5.30e-7 | 4.13 |
|  | 160 | 1.53e-7 | 9.03 | 2.19e-6 | 7.60 | 1.84e-8 | 4.01 | 3.17e-8 | 4.06 |
|  | 320 | 1.73e-9 | 6.47 | 5.53e-9 | 8.64 | 1.63e-9 | 3.50 | 3.06e-9 | 3.37 |
| $P^3$-RKDG | 40 | 7.39e-4 |  | 2.69e-3 |  | 1.01e-8 |  | 2.94e-8 |  |
|  | 80 | 9.80e-9 | 16.20 | 1.14e-7 | 14.53 | 1.22e-10 | 6.37 | 2.42e-10 | 6.93 |
|  | 160 | 1.74e-9 | 2.49 | 3.21e-8 | 1.82 | 5.89e-12 | 4.38 | 1.12e-11 | 4.44 |
|  | 320 | 5.36e-11 | 5.02 | 1.91e-9 | 4.07 | 3.35e-13 | 4.14 | 7.80e-13 | 3.84 |
| $P^2$-RKDG HWENO | 40 | 1.96e-3 |  | 5.48e-3 |  | 7.47e-5 |  | 2.32e-4 |  |
|  | 80 | 2.71e-4 | 2.85 | 1.53e-3 | 1.84 | 4.68e-6 | 3.99 | 1.96e-5 | 3.57 |
|  | 160 | 1.72e-5 | 3.97 | 1.98e-4 | 2.95 | 2.37e-7 | 4.30 | 8.29e-7 | 4.56 |
|  | 320 | 1.33e-8 | 10.34 | 5.55e-8 | 11.80 | 1.14e-8 | 4.37 | 3.64e-8 | 4.51 |
| $P^3$-RKDG HWENO | 40 | 1.21e-3 |  | 3.65e-3 |  | 2.68e-5 |  | 6.98e-5 |  |
|  | 80 | 5.94e-5 | 4.34 | 3.14e-4 | 3.54 | 5.12e-7 | 5.71 | 2.19e-6 | 4.99 |
|  | 160 | 2.81e-8 | 11.05 | 2.55e-7 | 10.26 | 2.40e-8 | 4.41 | 2.55e-7 | 3.09 |
|  | 320 | 2.25e-10 | 6.96 | 4.62e-9 | 5.78 | 6.35e-11 | 8.56 | 7.13e-10 | 8.48 |

**Figure Captions**

Figure 1 shows the results of the Riemann problem given by eqn. (5.1). The flow variables are : a) density, b) pressure, c) x-velocity, d) y-velocity and e) y-component of magnetic field.

Figure 2 shows the results of the Riemann problem given by eqn. (5.2). The flow variables are : a) density, b) pressure, c) x-velocity, d) y-velocity and e) y-component of magnetic field.

Figure 3 shows the results of the Riemann problem given by eqn. (5.3). The flow variables are : a) density, b) pressure, c) x-velocity, d) y-velocity, e) z-velocity, f) y-component of magnetic field and g) z-component of magnetic field.

Figure 4 shows the results of the Riemann problem given by eqn. (5.4). The flow variables are : a) density, b) pressure, c) x-velocity, d) y-velocity, e) z-velocity, f) y-component of magnetic field and g) z-component of magnetic field.

Figure 5 shows the results of the Riemann problem given by eqn. (5.5). The flow variables are : a) density, b) pressure, c) x-velocity, d) y-velocity, e) z-velocity, f) y-component of magnetic field and g) z-component of magnetic field.

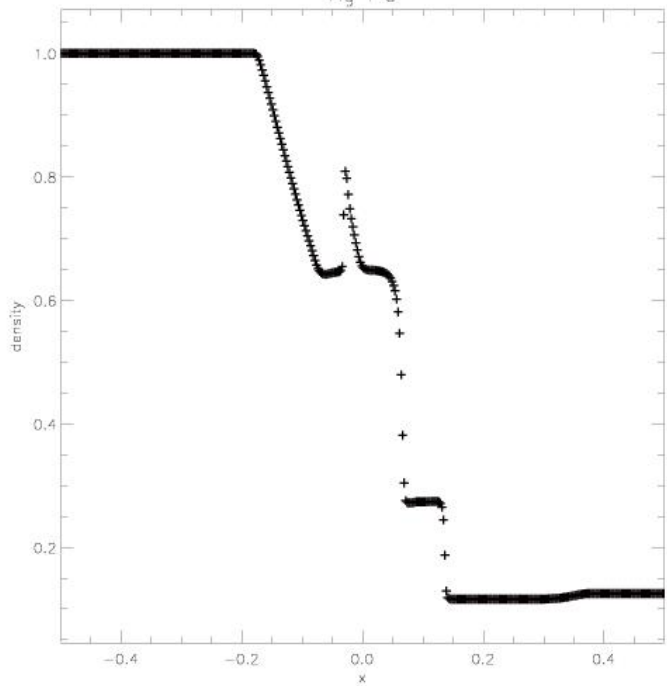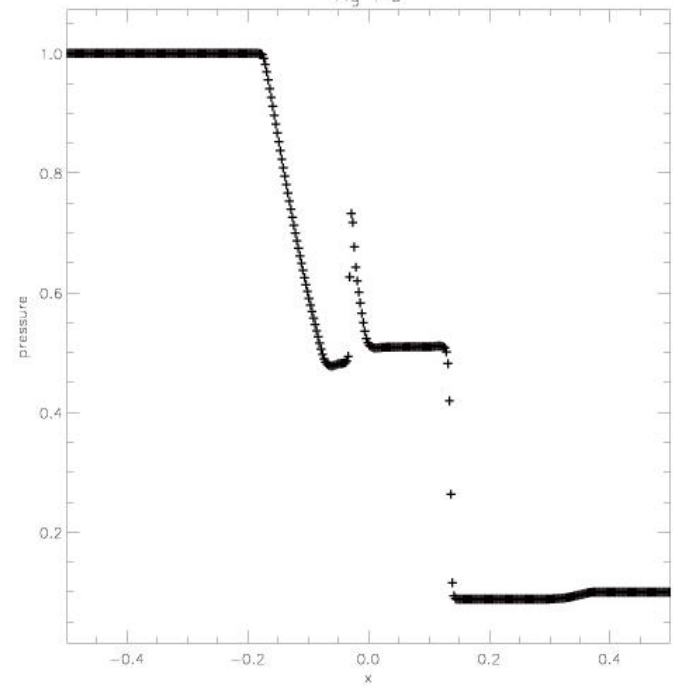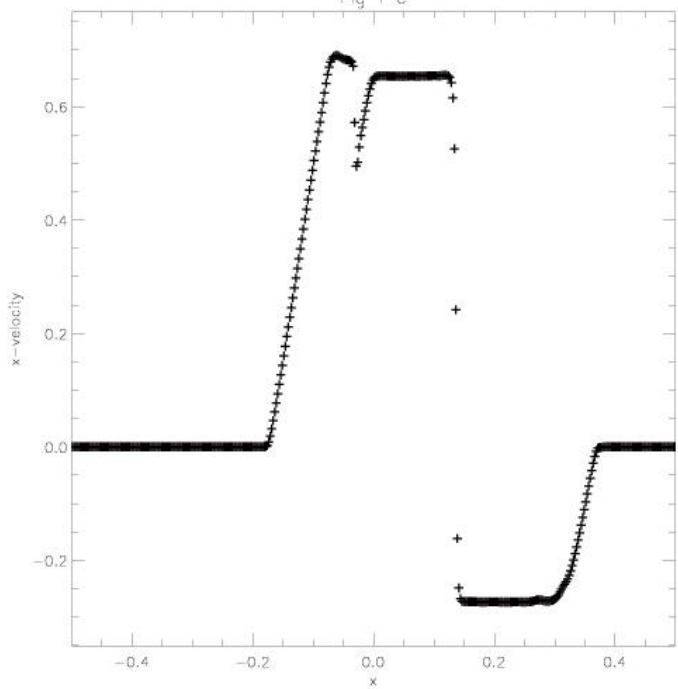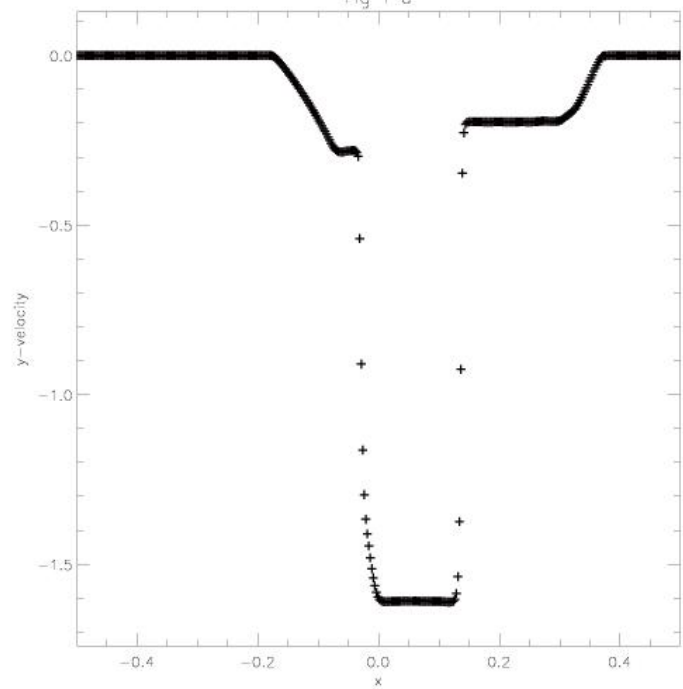

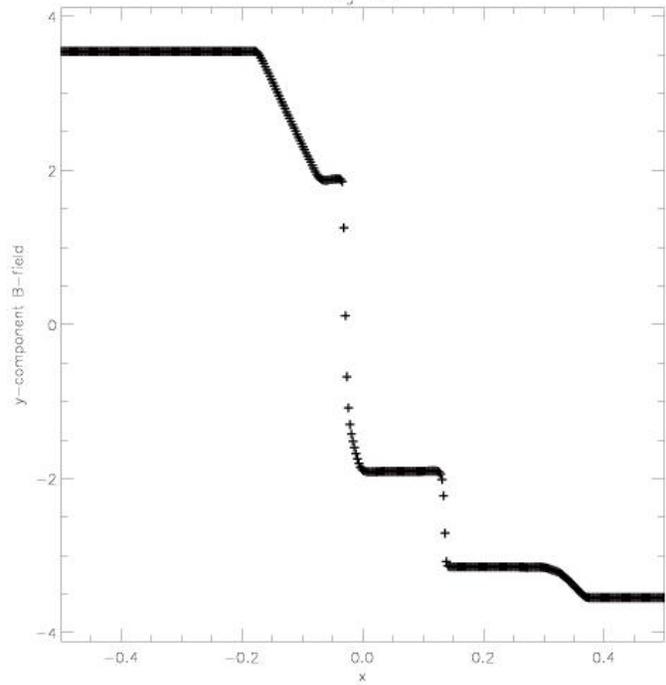

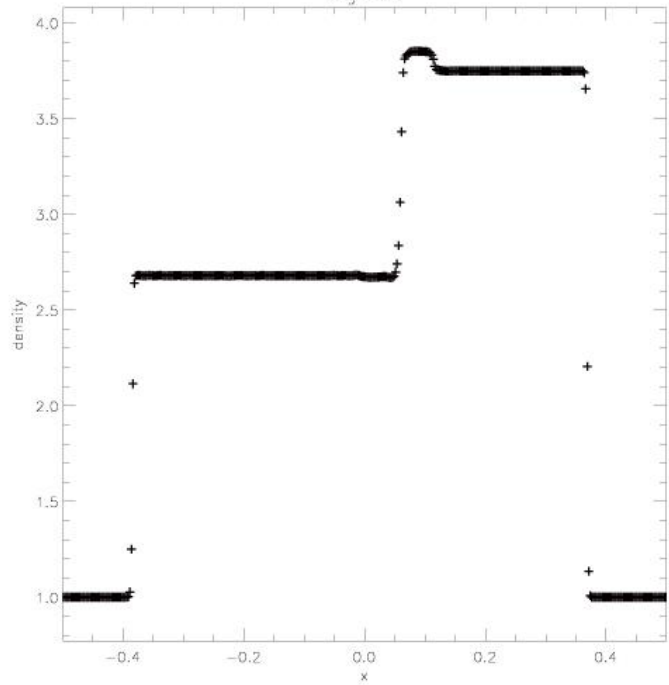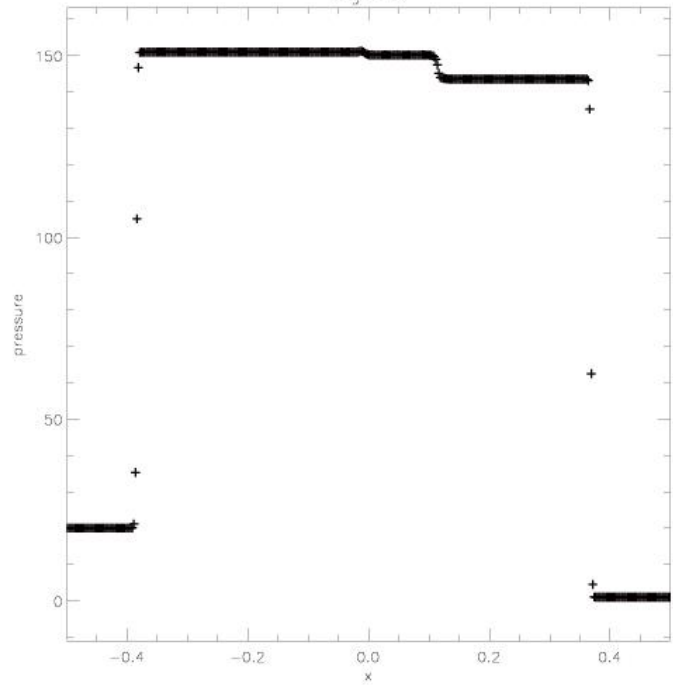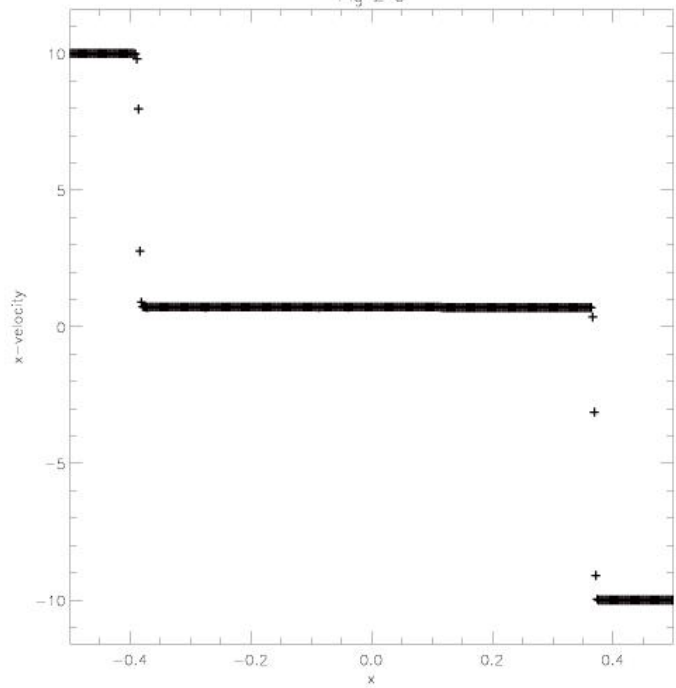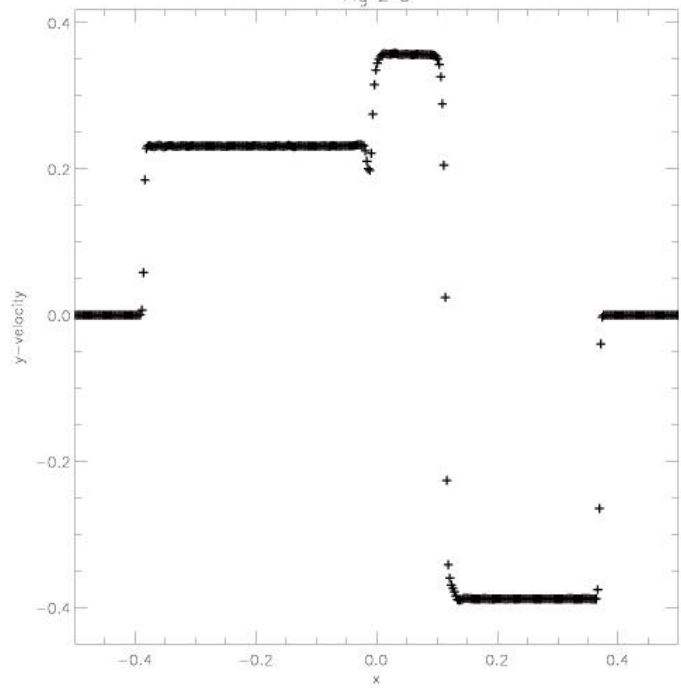

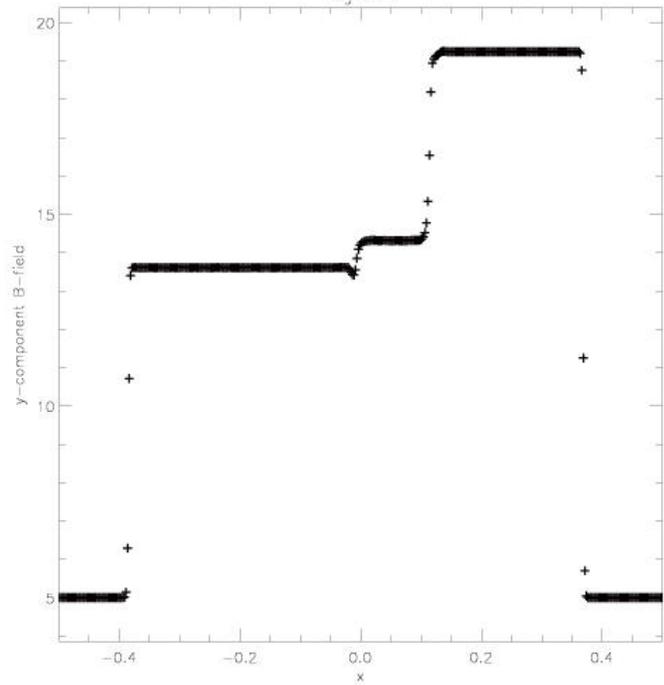

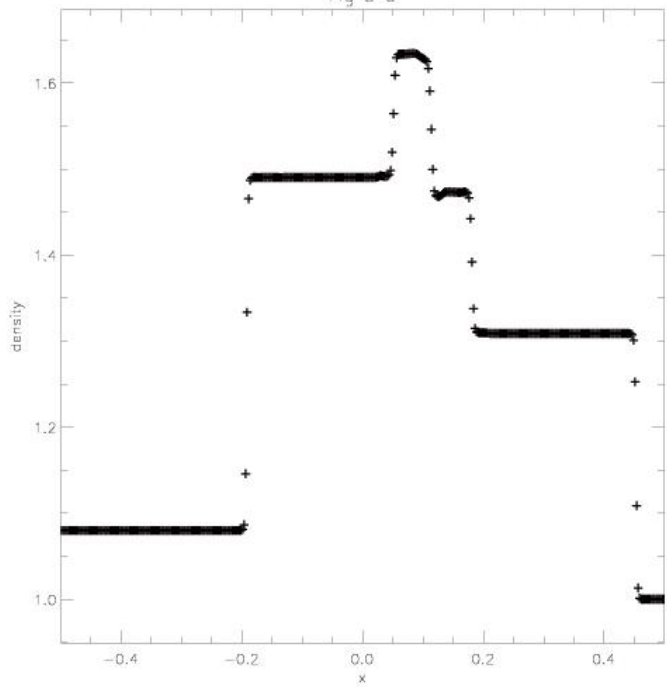
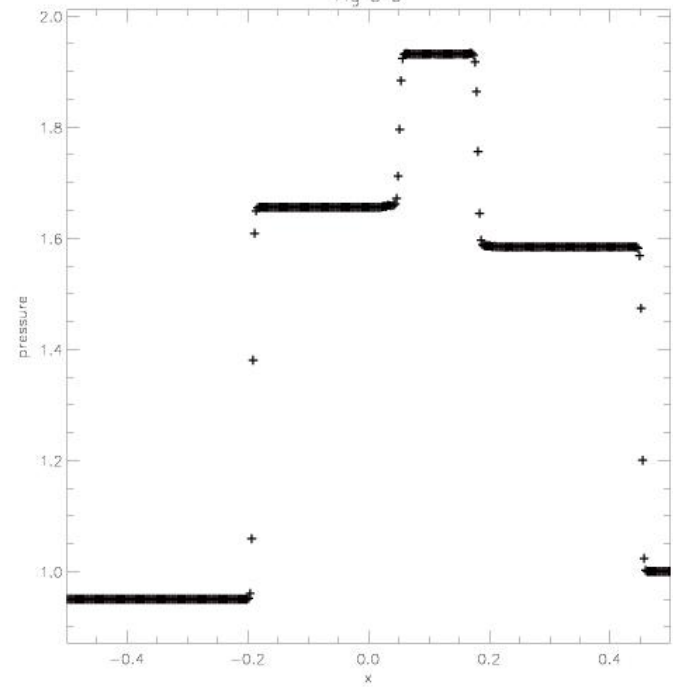
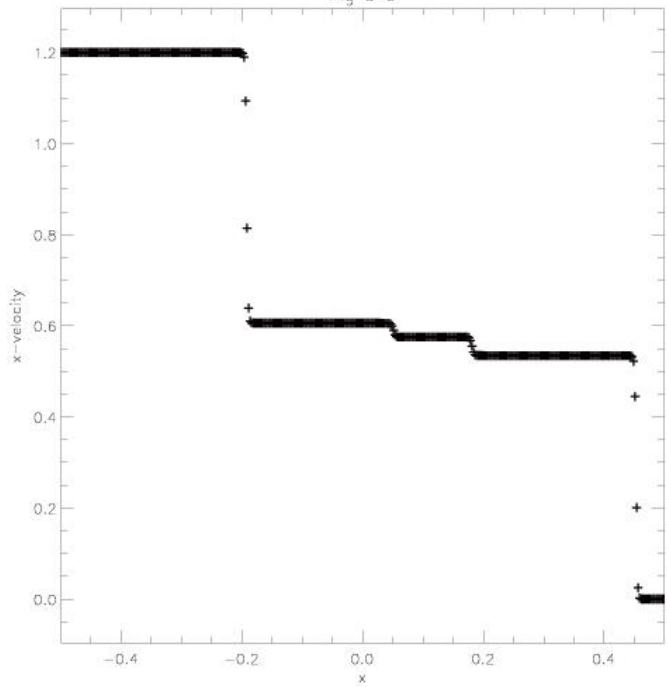
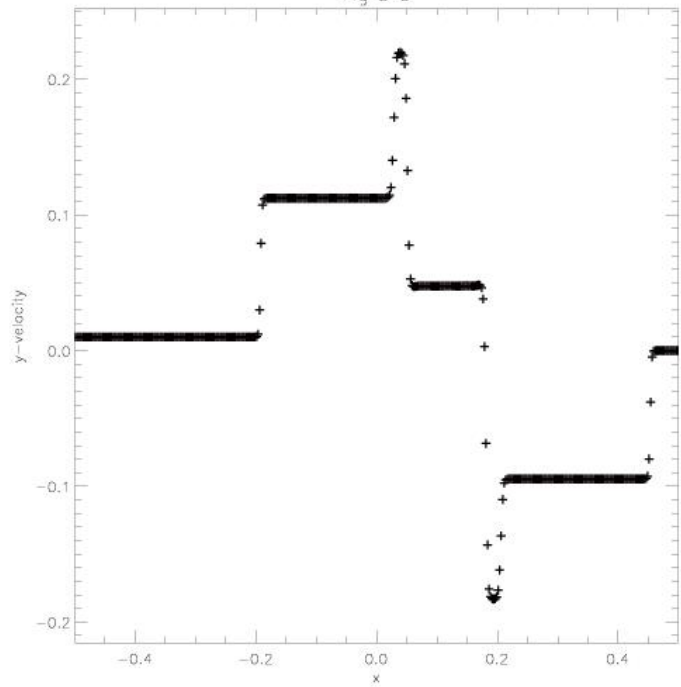

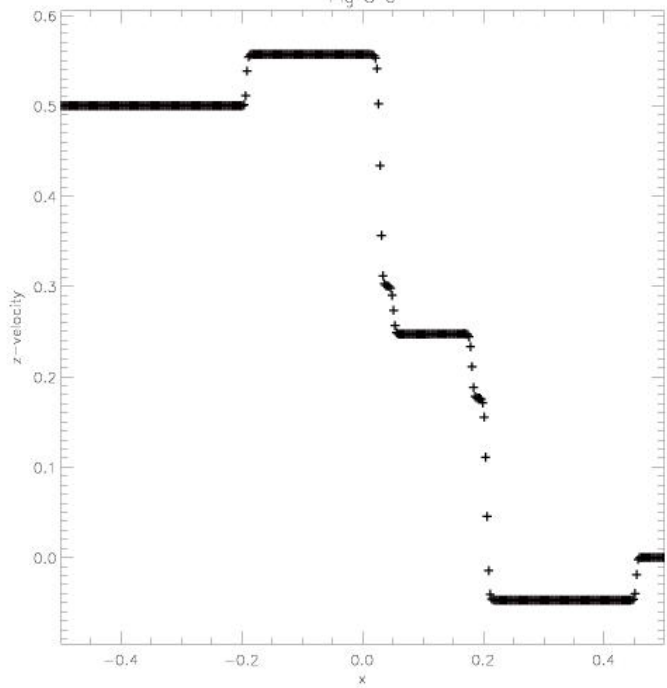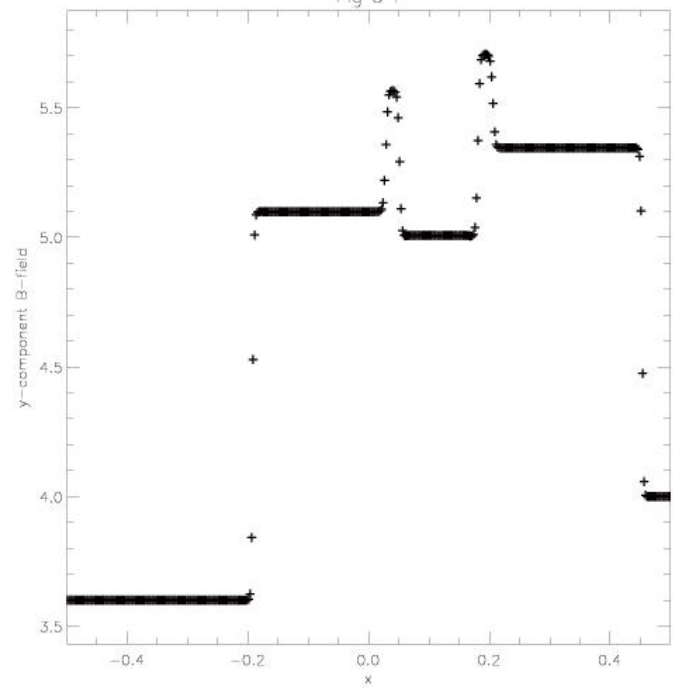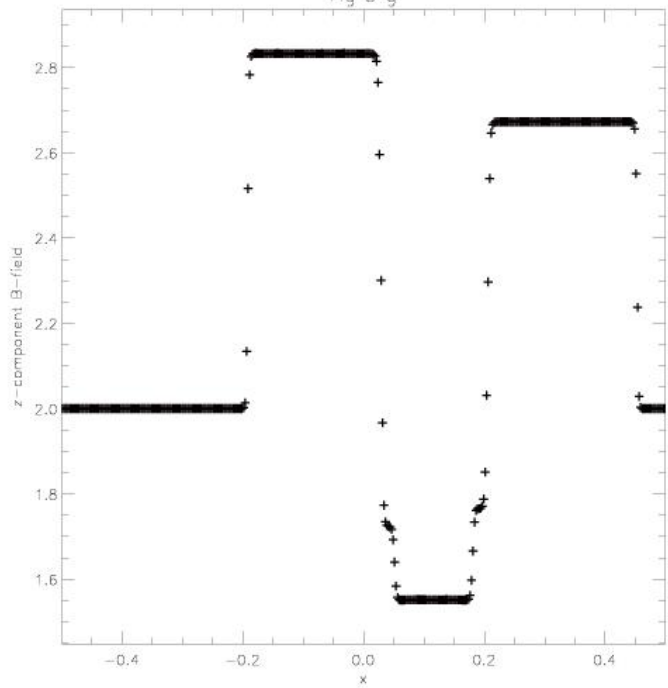

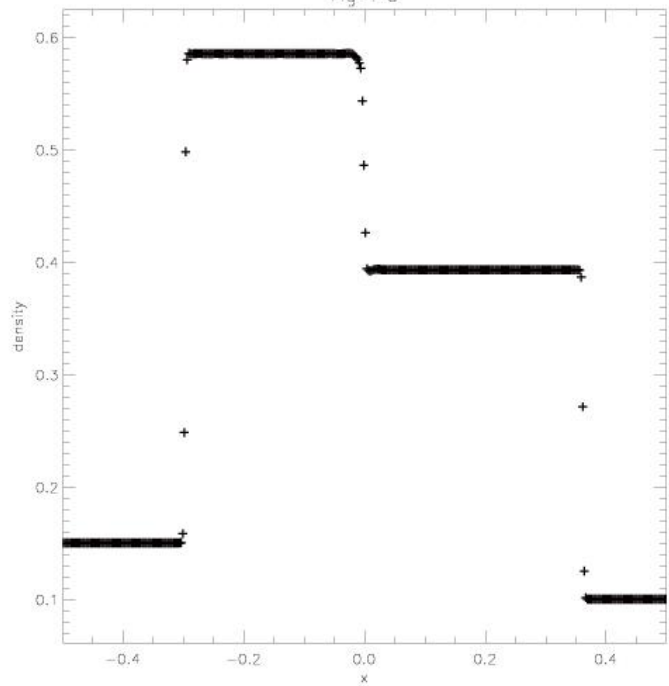 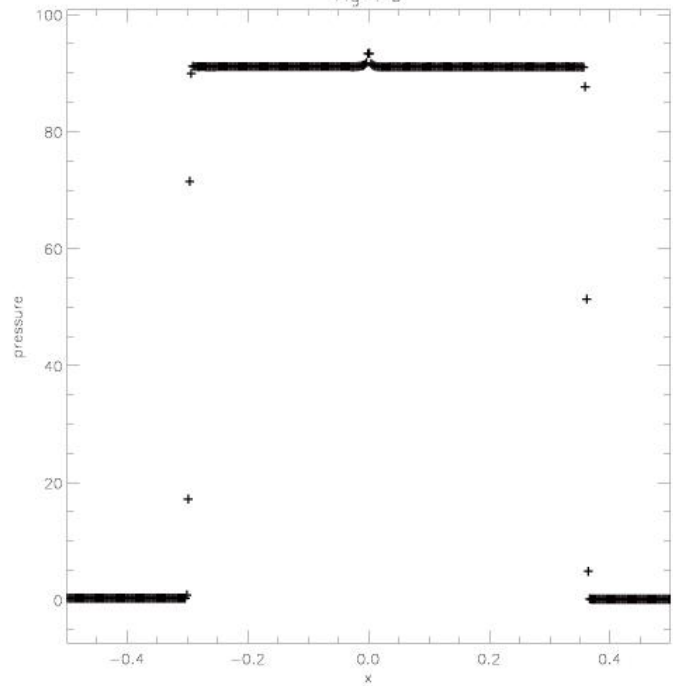 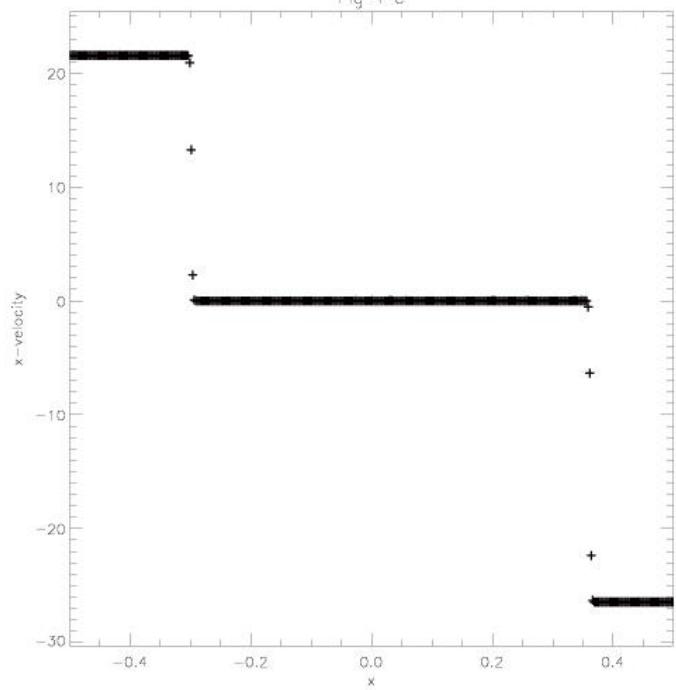 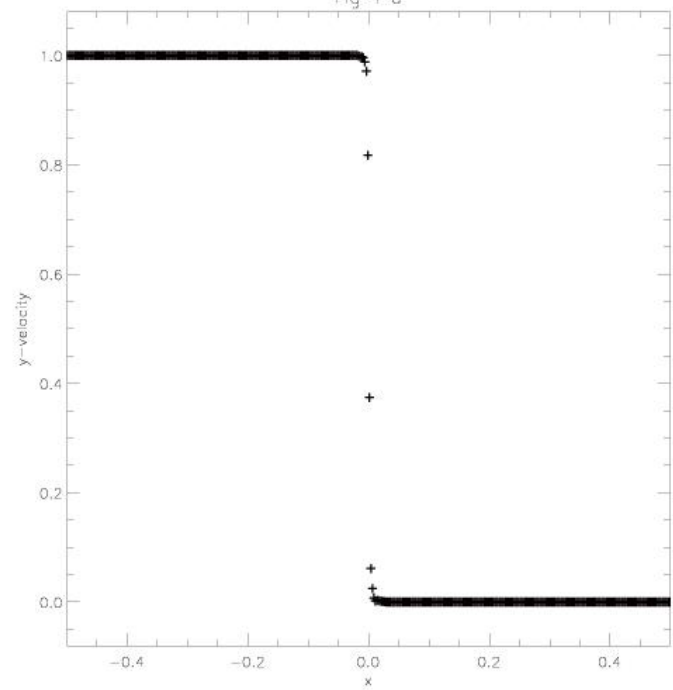

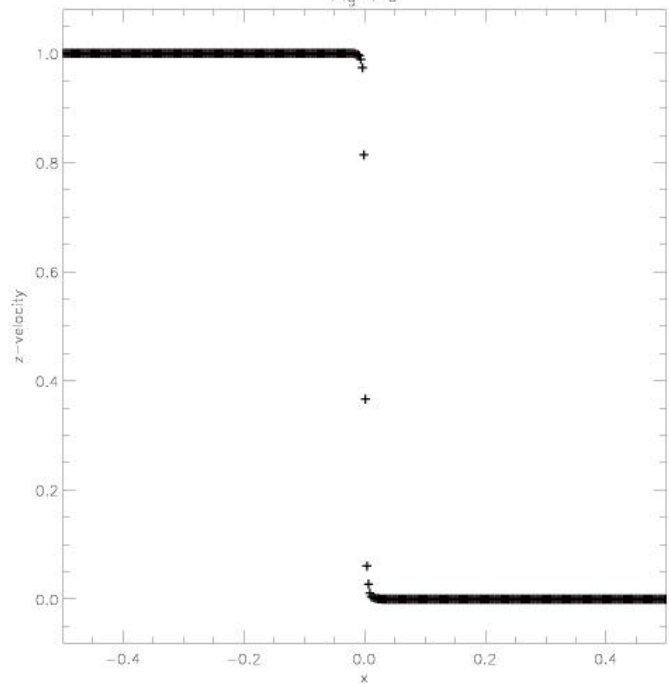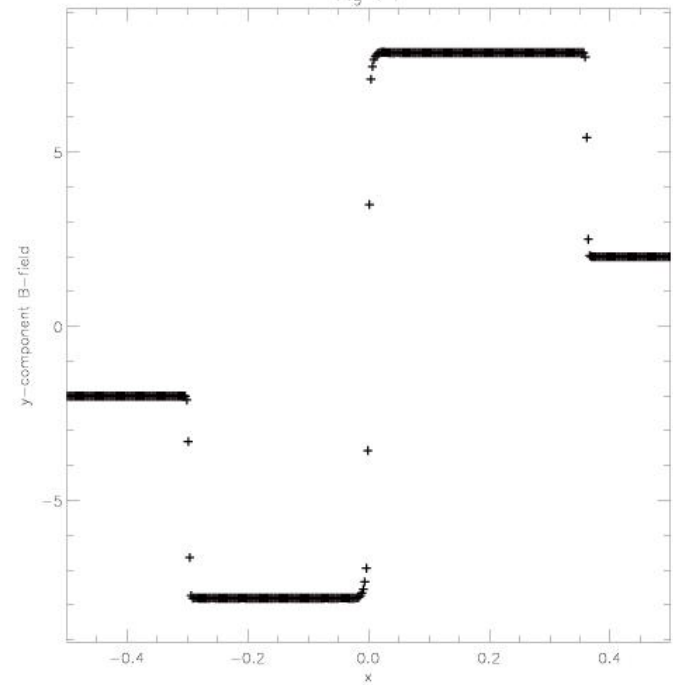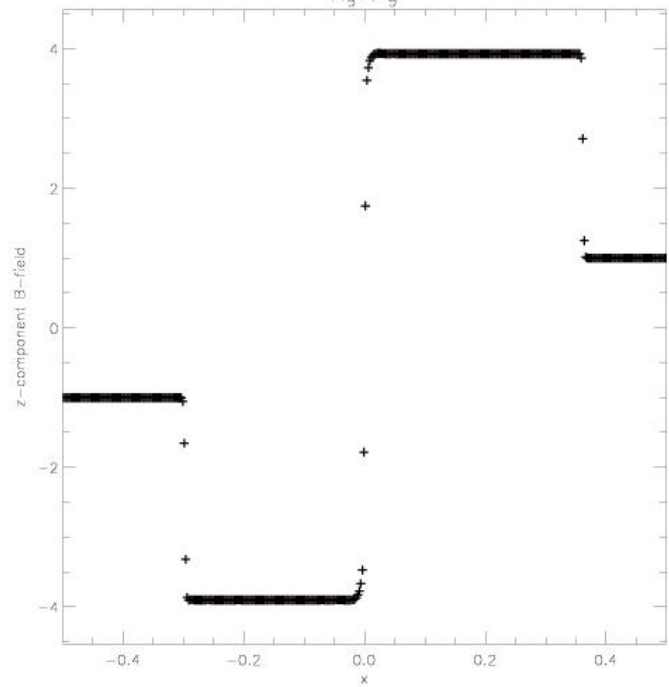

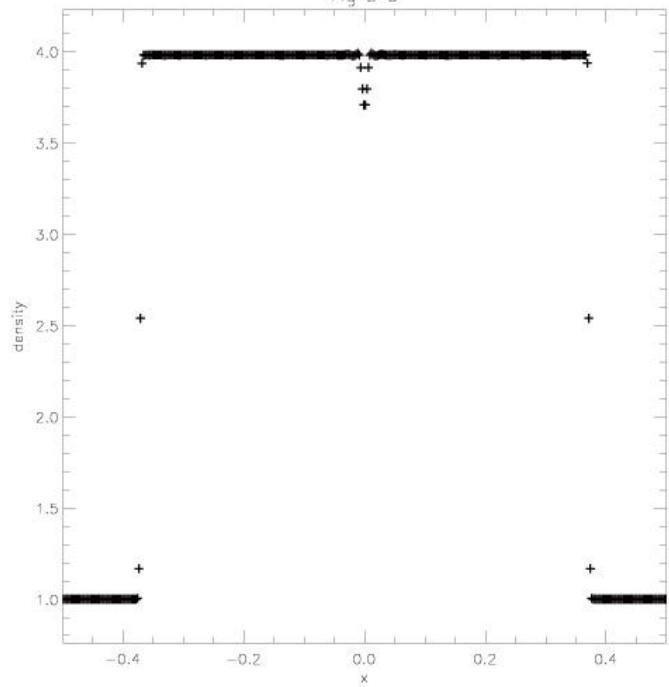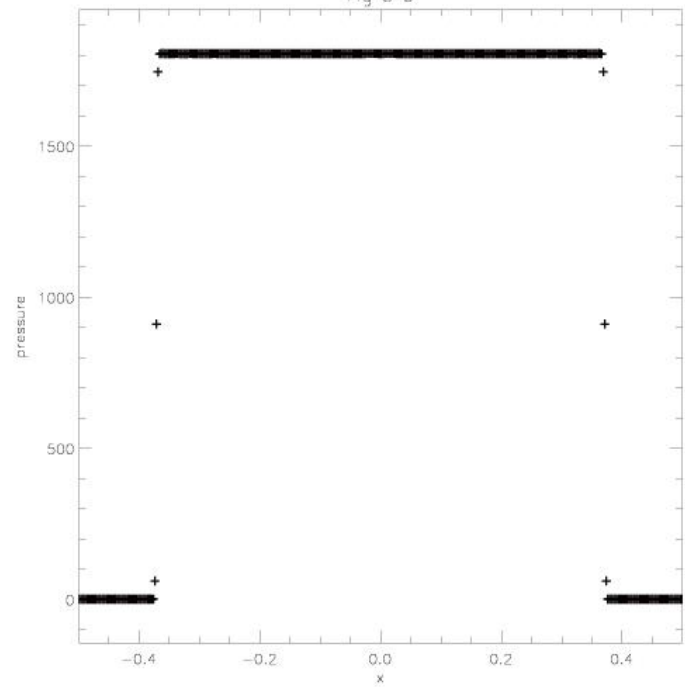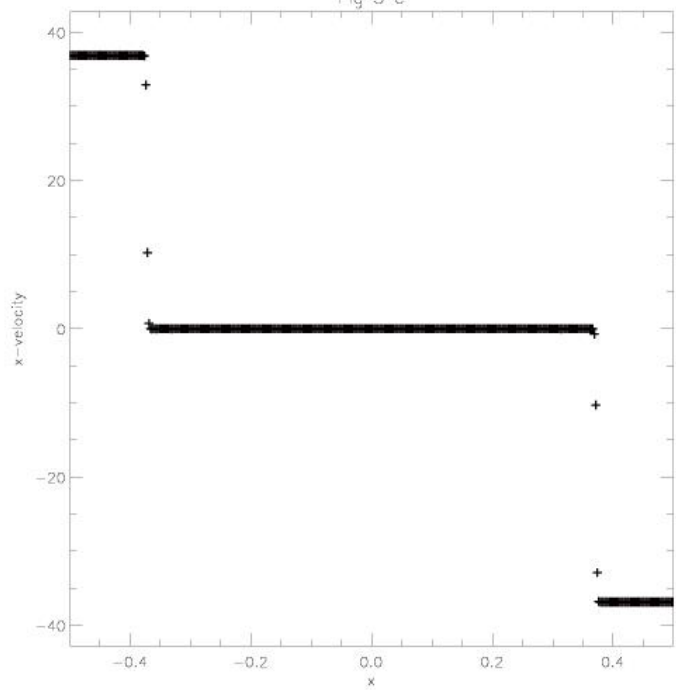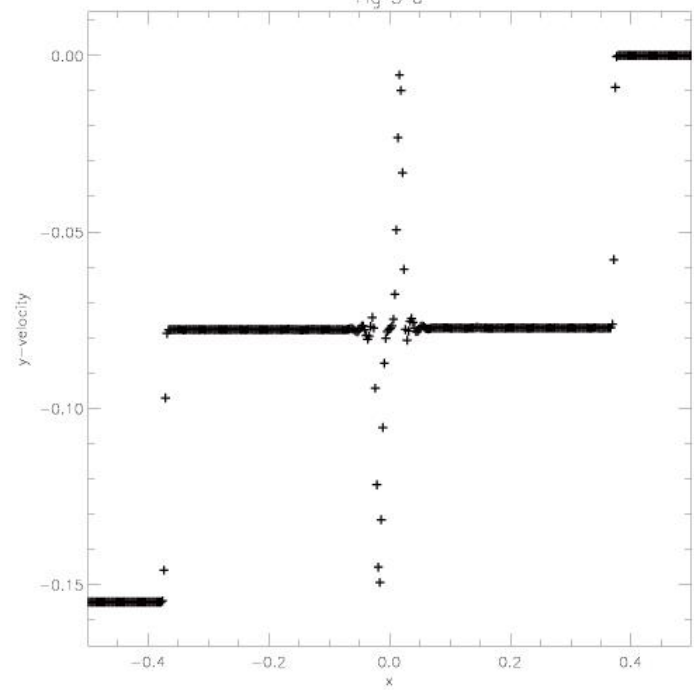

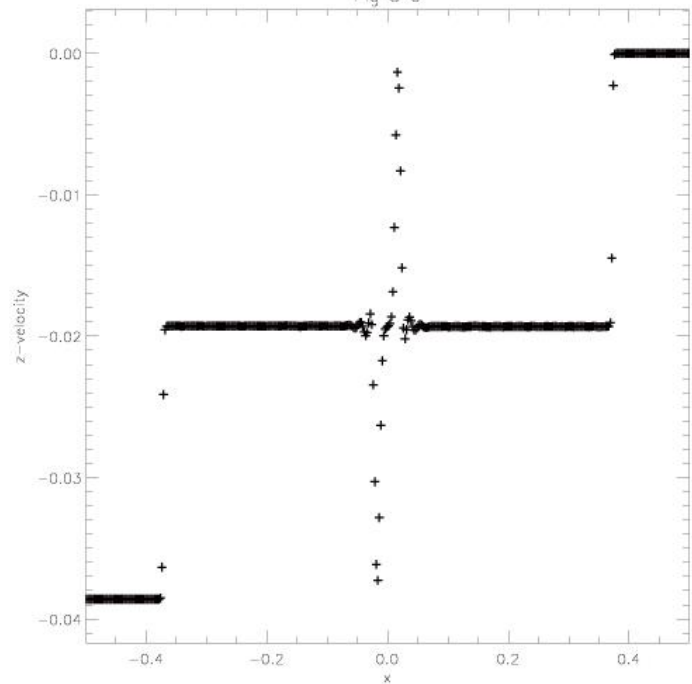

Fig 5 e

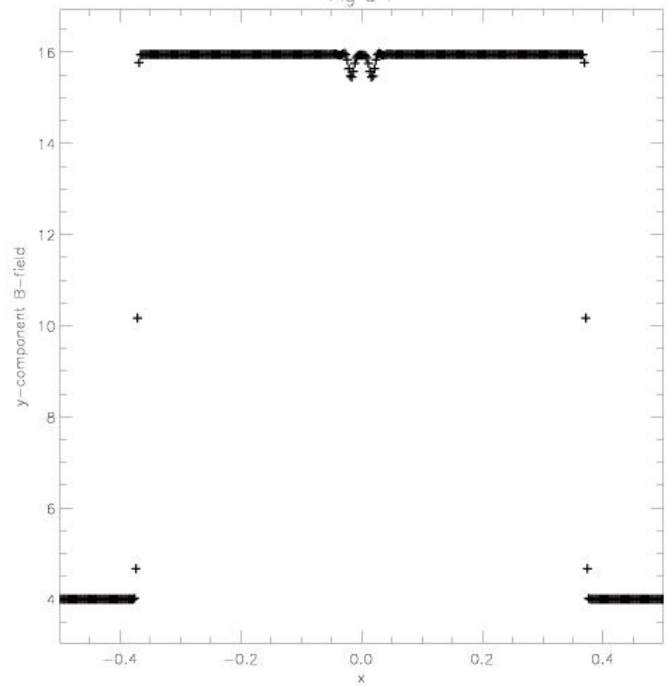

Fig 5 f

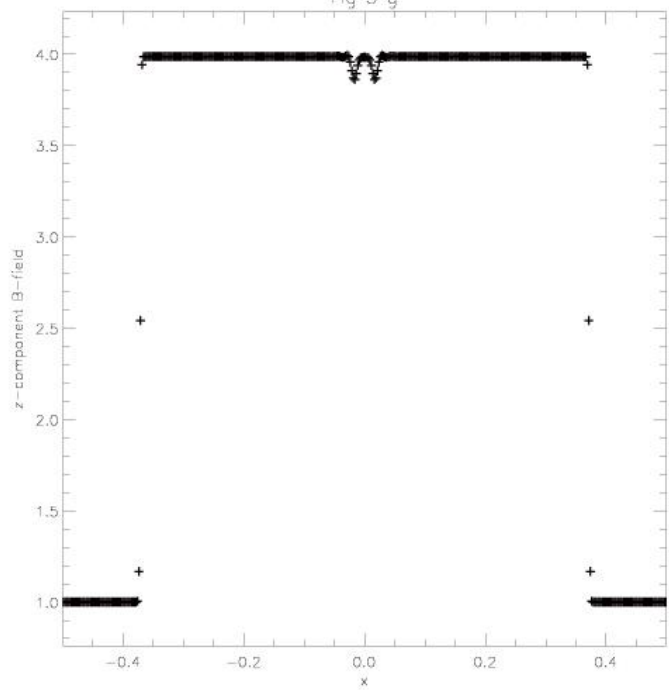

Fig 5 g